\documentclass[prl, twocolumn, letterpaper, showpacs]{revtex4}
\usepackage{graphicx}
\usepackage{amsmath}
\usepackage{bm}
\usepackage{hyperref}
\usepackage{verbatim}
\hypersetup{
    colorlinks,
    citecolor=red,
    filecolor=blue,
    linkcolor=blue,
    urlcolor=blue
}

\providecommand{\abs}[1]{\lvert#1\rvert}

\newcommand{\ket}[1]{\left| #1 \right>} 

\def\ket#1{{|#1\rangle}}

\begin{document}

\title{Optimal suppression of defect generation during a passage across a quantum critical point}
\author{Ning Wu\footnote{These two authors equally contributed to this work.}, Arun Nanduri$^*$, and Herschel Rabitz\footnote{hrabitz@princeton.edu}}
\affiliation{\it Department of Chemistry, Princeton University, Princeton, NJ 08544}

\begin{abstract}
 The dynamics of quantum phase transitions are inevitably accompanied by the formation of defects when crossing a quantum critical point. For a generic class of quantum critical systems, we solve the problem of minimizing the production of defects through the use of a gradient-based deterministic optimal control algorithm. By considering a finite size quantum Ising model with a tunable global transverse field, we show that an optimal power law quench of the transverse field across the Ising critical point works well at minimizing the number of defects, in spite of being drawn from a subset of quench profiles. These power law quenches are shown to be inherently robust against noise. The optimized defect density exhibits a transition at a critical ratio of the quench duration to the system size, which we argue coincides with the intrinsic speed limit for quantum evolution.
\end{abstract}


\maketitle

\emph{Introduction.}--Understanding the dynamics of quantum phase transitions~\cite{Sachdev, DziarmagaRev,gapless} is a topic of more than academic interest, as the production of defects during the crossing of a quantum critical point (QCP) hinders the ability to accurately prepare initial many-body states for use in quantum simulators~\cite{Cirac,Sorensen,Gericke} and quantum computation~\cite{Farhi,Nielsen}. However, the exact preparation of a many-body ground state is a highly nontrivial problem due to the difficulty of evaluating the real time dynamics of interacting many-body systems. Achieving this goal by naively tuning a parameter of the Hamiltonian is only possible if the process is carried out on impractically long time scales.
\par Recently, numerous analyses for achieving an optimal passage through a QCP have emerged~\cite{PRA2002,Campo,nonHermitian, Campo1, Damski, LMG, Sau, PRE,Sen, Jopt,EPreview,Bason,Sen2007,Barankov, Power, Doria, CanevaPRA,Rahmani, sLloyd}. In general, the methods used in these studies can be classified into two categories (see Ref.~\cite{EPreview} for a recent review): I. The transitionless quantum driving proposed in Ref.~\cite{Campo} and developed in Refs.~\cite{Campo1, Damski,LMG, Sau,Jopt}. This type of quantum control requires the addition of generally complex interaction terms to the system. Although such protocols have been implemented experimentally in few-body systems~\cite{Bason}, their application in realistic many-body systems remains a challenge. II. Quantum optimal control methods, a natural idea for tuning terms already present in the original Hamiltonian. Based on the universal dynamics of phase transitions, Refs.~\cite{Sen2007,Barankov,Power} show that the use of power law protocols to traverse a QCP enables suppression of defect formation in the thermodynamic limit. However, only recently have there been a few attempts to apply optimal control techniques to many-body dynamics~\cite{Doria, CanevaPRA, Rahmani, sLloyd}.
In spite of these developments, there are no systematic algorithms to deal with the dynamical control of general many-body systems.

In this work, we propose using a gradient-based optimal control algorithm~\cite{Brif} to minimize excitations when crossing a QCP of a generic class of quantum critical systems. These systems can be mapped into free-fermion models, for which the gradient of the control observable with respect to the control field can be elegantly calculated analytically. Unlike the stochastic algorithms used in Ref.~\cite{Doria} and Ref.~\cite{Rahmani}, our method is deterministic and physically transparent. It can suppress excitations monotonically by choosing a suitable initial control. We stress that we seek to tune only parameters present in the original Hamiltonian. As an illustration, we show that the kink densities formed across a paramagnetic to ferromagnetic phase transition in the quantum Ising model can be greatly minimized by an optimal power law resulting from our method.

Furthermore, although it has been shown that optimal control can drive few-body systems at rates up to the quantum speed limit (QSL)~\cite{Bason, Caneva}, the relation between optimal control and the QSL remains to be understood in many-body dynamics. Intriguingly, we show that our result is indeed consistent with the QSL, and illustrate an intrinsic connection between the QSL and gradient-based optimal control. The power law nature of our optimal protocol renders it highly robust against noise. Finally, we comment on possible experimental platforms in which our protocol can be tested.

\emph{Model and methodology}.--We consider a closed quantum system described by a time-dependent Hamiltonian $H[g(t)]$, with instantaneous ground state $|G[g(t)]\rangle$, where the control field $g(t)$ can be tuned arbitrarily with certain constraints. We focus on a quench crossing a QCP within a finite time interval $t\in[-T,T]$. Fixing the endpoints $g_i=g(-T)$ and $g_f=g(T)$ of the control field, we aim to minimize (or maximize) the final expectation value $O(T)\equiv\langle \phi(T)| \hat{O} |\phi(T)\rangle$ of some controlled general observable $\hat{O}$ following the quench. Here $|\phi(t)\rangle=U(t,-T)|G(g_i)\rangle$ with the evolution operator $U(t,-T)=\mathcal{T}e^{-i\int^t_{-T}d\tau H[g(\tau)]}$. We focus on a family of $d$-dimensional free-fermion Hamiltonians that can be written as a summation over independent $\mathbf{k}$-modes in momentum space:
\begin{eqnarray}\label{Hkmode}
H=\sum_\mathbf{k}\psi^\dag_\mathbf{k}[\mathbf{d}_\mathbf{k}(g(t))\cdot\vec{\sigma}_\mathbf{k}]\psi_{\mathbf{k}},
\end{eqnarray}
where $\vec{\sigma}_\mathbf{k}=(\sigma^x_\mathbf{k},\sigma^y_\mathbf{k},\sigma^z_\mathbf{k})$ are Pauli matrices acting on the mode $\mathbf{k}$ and $\psi_\mathbf{k}=(a_\mathbf{k},b_\mathbf{k})^T$ are fermionic operators. The function $\mathbf{d}_\mathbf{k}(g )=(d^x_\mathbf{k}(g),d^y_\mathbf{k}(g),d^z_\mathbf{k}(g))$ is determined by the specific model. Eq. (\ref{Hkmode}) can represent a variety of systems exhibiting quantum phase transitions, e.g., the anisotropic $XY$ model in $d=1$~\cite{Lieb1962} and the Kitaev model in $d=1$ and 2~\cite{Sen2007,Lee2007,Wu2012}.
Eq.~(\ref{Hkmode}) can be diagonalized as $H=\sum_k\varepsilon_\mathbf{k}(\Psi^\dag_\mathbf{k}\Psi_\mathbf{k}-1)$ with dispersion $\varepsilon_\mathbf{k}=|\mathbf{d}_\mathbf{k}(g)|$ and $\Psi_\mathbf{k}=(A_\mathbf{k}, B_\mathbf{k})^T$, where $A_\mathbf{k}$ and $B_\mathbf{k}$ are suitable fermionic operators. The ground state is defined by $A_\mathbf{k}|G(g)\rangle=B_\mathbf{k}|G(g)\rangle=0$, $\forall~\mathbf{k}$.
Here, both $|G[g(t)]\rangle$ and the evolution operator are factorable: $|G [g(t)]\rangle=\prod_\mathbf{k}|G_\mathbf{k}[g(t)]\rangle$, and $U(t,-T)=\prod_kU_\mathbf{k}(t,-T)$. We assume that the global control field $g(t)$ enters the Hamiltonian via the term $\sim g(t)\hat{Y}$, where $\hat{Y}$ is some \emph{time-independent} operator. We further assume that both $\hat{Y}$ and the
controlled observable $\hat{O}$ can be written as a summation over \emph{even} operators of the independent $\mathbf{k}$-modes, $\hat{Y}=\sum_\mathbf{k} \hat{F}_\mathbf{k},~\hat{O}=\sum_\mathbf{k} \hat{O}_\mathbf{k}$.
A wide variety of observables satisfy this condition, and interesting examples include the kink density~\cite{Dziarmaga} and state fidelity in the transverse Ising chain, the survival probability of edge states in p-wave superconductors~\cite{MF}, etc. The evenness of $\hat{F}_\mathbf{k}$ ensures that the evolved state $|\phi_\mathbf{k}(t)\rangle=U_\mathbf{k}(t,-T)|G_\mathbf{k}(g_i)\rangle$ conserves the parity of the initial state and should be a linear combination of $|G_\mathbf{k}[g(t)]\rangle$ and $|\bar{G}_\mathbf{k}[g(t)]\rangle=A^\dag_\mathbf{k}(t)B^\dag_\mathbf{k}(t)|G_\mathbf{k}[g(t)]\rangle$. The gradient $\frac{\delta O( T)}{\delta g( t)}$, a key quantity in the optimal control algorithm we employ, then can be calculated as (see~\cite{SM} for the derivation)
\begin{eqnarray}\label{gradO}
\frac{\delta O( T)}{\delta g( t)}&=&2\Im\langle G(g_i)| \hat{O} (T)\hat{Y}(t)|G(g_i)\rangle\nonumber\\
&=&2\Im\sum _{\mathbf{k}}\langle \phi_\mathbf{k}(T)| \hat{O}_\mathbf{k}  |\bar{\phi}_\mathbf{k}(T)\rangle \langle \bar{\phi}_\mathbf{k}(t)| \hat{F}_{\mathbf{k} } |\phi_\mathbf{k}(t)\rangle,
\end{eqnarray}
where $\hat{X}(t)\equiv U^\dag(t,-T)\hat{X}U(t,-T)$ for operator $\hat{X}$ and $|\bar{\phi}_\mathbf{k}(t)\rangle=U_\mathbf{k}(t,-T)|\bar{G}_\mathbf{k}(g_i)\rangle$. We note that in general, the concise expression above does not hold for nonintegrable interacting many-body systems. Instead, in order to evaluate $|\phi(t)\rangle$ in the gradient, one has to pursue advanced numerical techniques~\cite{tDMRG,tDMFT} for treating general many-body dynamics.

\emph{Application to the quantum Ising model}.--We now focus on the quantum Ising chain with $N$ spins in a controllable transverse magnetic field $g(t)$,
\begin{eqnarray}\label{Isingchain}
H(t)=-\sum^N_{j=1}[\sigma^x_j\sigma^x_{j+1}+g(t)\sigma^z_j].
\end{eqnarray}
For simplicity, we consider periodic boundary conditions $\vec{\sigma}_{N+1}=\vec{\sigma}_1$ and even $N$. This model exhibits a quantum phase transition at $g_c=1$ between the ferromagnetic phase for $0\leq g<1$ and paramagnetic phase for $g>1$. The Jordan-Wigner transformation $\sigma^z_j=2c^\dag_jc_j-1$, $(\sigma^x_j+i\sigma^y_j)/2= c_je^{i\pi\sum^{j-1}_{l=1}c^\dag_lc_l}$, followed by a Fourier transform $c_j=e^{i\frac{\pi}{4}}/\sqrt{N}\sum_{k}e^{ikj}c_k$ 
maps Eq.~(\ref{Isingchain}) into a free model $H(t)=\sum_{k>0}\Lambda_k(t)(\eta^\dag_k\eta_k+\eta^\dag_{-k}\eta_{-k}-1)$, where the spectra $\Lambda_k(t)=2\sqrt{[g(t)+\cos k]^2+\sin^2k}$ and 
the quasiparticle operator $\eta_k=\cos\theta_kc_k-\sin\theta_k c^\dag_{-k}$ with $\tan2\theta_k=-\sin k/(g+\cos k)$.
The instantaneous ground state of mode $k$ can be written in the basis $\{|0_k\rangle,c^\dag_kc^\dag_{-k}|0_k\rangle\}$ as $|G _k[g(t)]\rangle=(\cos\theta_k[g(t)], \sin\theta_k[g(t)] )^T$. 

\par We will consider a passage from the paramagnetic to the ferromagnetic phase, and attempt to find an optimized pulse which produces the lowest number of defects (or kinks) $\hat{D}=\frac{1}{2}\sum^N_{j=1}(1-\sigma^x_j\sigma^x_{j+1})=2\sum _{k>0} \hat{P}_k$, with
\begin{eqnarray}\label{D}
  \hat{P}_k&=&\frac{1}{2}[1-\psi^\dag_k(\cos k \sigma^z_k+\sin k\sigma^x_k)\psi_k],
\end{eqnarray}
whose expectation value in the state $|\bar{\phi}_k(T)\rangle$ gives the excitation probability $P_k(T)=|\langle G_k(g_f)|\bar{\phi}_k(T)|^2$ of the pair of modes $(k,-k)$ \cite{Dziarmaga}. Here $\psi_k=(c_k,c^\dag_{-k})^T$. By Eq.~(\ref{gradO}), the gradient reads
\begin{eqnarray}\label{gradD}
\frac{\delta D(T)}{\delta g(t)}&=&-4\Im\sum _{k>0}\langle\bar{\phi}_k(t)|\sigma^z_k|\phi_k(t)\rangle\nonumber\\
&&\langle \phi_k(T)| \sin k\sigma^x_k+ \cos k(\sigma^x_k+i\sigma^y_k) | \bar{\phi}_k(T)\rangle.
\end{eqnarray}

\par The simplest quench profile is linear, for which the resulting density of defects obeys the Kibble-Zurek scaling and a natural time scale $T_{\rm{ad}}\propto N^2$ exists separating the adiabatic $(T>T_{\rm{ad}})$ and non-adiabatic $(T<T_{\rm{ad}})$ regimes~\cite{Dziarmaga}. Although such a crossover time is not well-defined for time-dependent profiles $g(t)$, we will focus on the regime $T<T_{\rm{ad}}$.
As a simple extension to the linear quench, a power law profile has been used~\cite{Barankov} to optimally cross a QCP in infinite critical systems. The physical arguments in Ref.~\cite{Barankov} indicate that an optimal power $r^*$ should also exist for \emph{finite} size systems. As we illustrate later, in spite of comprising a subset of quench profiles, power law quenches work well in robustly minimizing the defect density.

\par We consider a symmetric quench from $g_i=2$ to $g_f=0$ over the time interval $[-T,T]$ which we discretize into $10^4$ points. Then, the power-law quench to be optimized is of the form $g(r,t)=1-\abs{\frac{t}{T}}^r\mathrm{sgn}(t)$. To find the optimal power, we smoothly vary $r$ from an initial guess by introducing a parameter $s$ such that $r\to r(s)$, $s\geq 0$. This continuous variable is used to parameterize the trajectory of the control field $g(r(s),t)$ as the gradient search progresses~\cite{Brif}. We require the defect density at time $T$ to decrease as $s$ is increased,
\begin{equation}\label{optimizer1}
\frac{dD(r(s),T)}{ds}=\frac{dr(s)}{ds}\int^T_{-T} dt\frac{\delta D(r(s),T)}{\delta g(r(s),t)}\frac{\partial g(r(s),t)}{\partial r(s)}<0,
\end{equation}
which can be fulfilled by updating $r$ according to
\begin{eqnarray}\label{optimizer2}
\frac{dr(s)}{ds}=\int^T_{-T} dt\; \abs{\frac{t}{T}}^r \mathrm{sgn}(t)\; \ln \abs{\frac{t}{T}}\frac{\delta D(r(s),T)}{\delta g(r(s),t)}.
\end{eqnarray}
Here, the gradient $\frac{\delta D(r(s),T)}{\delta g(r(s),t)}$ is given by Eq.~(\ref{gradD}). The algorithm terminates when $\frac{dD(r(s),T)}{ds}=0$ to acceptable precision.
\par
\begin{figure}[h!]
  \includegraphics[width=.5\textwidth]{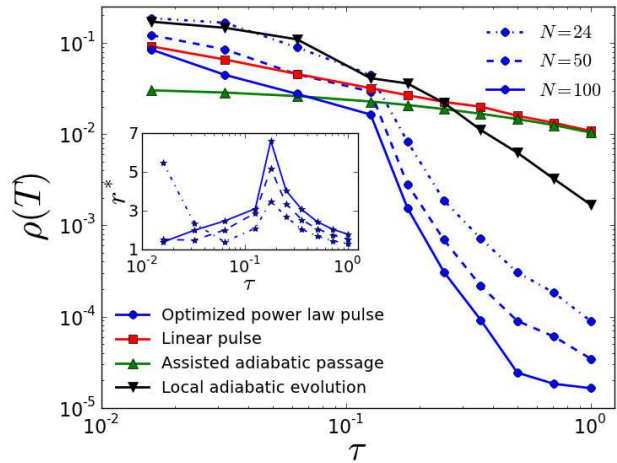}
  \caption{\label{defects} The defect density $\rho(T)$ (blue circles) obtained by applying a power law quench with a power $r^*$ obtained through the gradient algorithm, compared with the results for a linear quench~\cite{Dziarmaga} (red squares), the truncated assisted driving with $M=10$~\cite{Campo} (green triangles), and the local adiabatic evolution~\cite{PRA2002} (black triangles). All the solid lines are plotted using $N=100$.
    The inset displays the optimal power $r^*$ found by the gradient algorithm for different $\tau$ and system sizes. For $\tau>\tau_c$, $r^*$ takes on a uniquely defined large value that increases with $N$ and decreases with $\tau$. When $\tau<\tau_c$, the value for $r^*$ found by the gradient algorithm is not unique due to traps in the control landscape, and this is reflected in the erratic behavior of $r^*(\tau)$.}
\end{figure}

 In Fig.~\ref{defects}, we present the optimized final defect density $\rho(T)=D(T)/N$ as a function of the scaled time $\tau=T/N$, resulting from applying a power law pulse with an optimized power $r=r^*$ determined by the gradient algorithm to the Ising model with $N=24$, 50, and 100.
We compare our results for $N=100$ with those resulting from (i) the linear quench $g_l(t)=1-\frac{t}{T}$~\cite{Dziarmaga}, (ii) the local adiabatic evolution $\arctan\frac{g_a(t)+\cos k_N}{\sin k_N}= \frac{1}{2} [(1-\frac{t}{T})\arctan\frac{2+\cos k_N}{\sin k_N}+(1+\frac{t}{T})\arctan\frac{ \cos k_N}{\sin k_N}]$~\cite{PRA2002}, where $k_N=\pi-\frac{\pi}{N}$ is the lowest mode determining the first excited state,
and also with (iii) the transitionless quantum driving using a linear quench $g_l(t)$ and $M=10$~\cite{Campo}.

It is apparent that after a critical quench time $T_c$, the optimized defect density drops sharply. The systems with different numbers of spins seem to exhibit this drop at a constant value of $0.126<T_c/N=\tau_c<0.178$, suggesting that $T_c\propto N$, in contrast with the adiabatic time scale
$T_{\rm{ad}}\propto N^2$. For $\tau<\tau_c$, our algorithm only modestly outperforms the linear quench, and the truncated driving procedure yields better results for relatively small values of $M\lesssim 10$. On the other hand, for $\tau>\tau_c$, the optimized defect density drops sharply with increasing $\tau$, significantly outperforming the other three procedures by several orders of magnitude. Intriguingly, the behavior of the one-dimensional control landscape $D(r)$ changes abruptly close to $\tau_c$ (not shown). When $\tau<\tau_c$, there are numerous local minima in the landscape, and the final power $r^*$ found by the gradient algorithm depends on the initial value of $r$ chosen. Since there is no \emph{a priori} way to infer which local minimum is the global one, the performance of the gradient algorithm, which halts as soon as a local minimum is found, is greatly inhibited by the presence of these ``traps''. On the other hand, for $\tau>\tau_c$, we have observed that a unique global minimum appears in the control landscape $D(r)$. That is, no matter what initial value for $r$ is chosen, the gradient algorithm is always able to find the globally optimal power $r^*$. 

\begin{figure}[h!]
  \includegraphics[width=.5\textwidth]{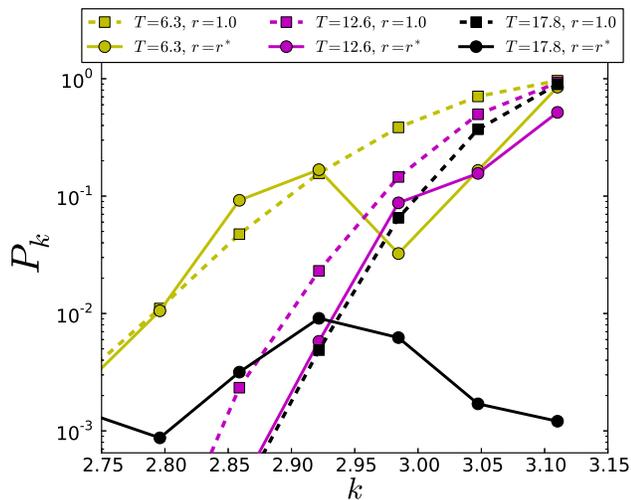}
  \caption{\label{Pk}The excitation probability $P_k$ for different quench durations $T$ with $N=100$. Shown are the results for both a linear quench with $r=1.0$ (dashed lines) and the optimal power law quench with $r=r^*$ (solid lines). For $T>T_c$, optimization of the control pulse causes $P_{k_N}$ to drop sharply; this occurs concurrently with the drastic improvement in $\rho(T)$.}
\end{figure}

\emph{Relation to the QSL}.--The existence of a global optimum is one factor contributing to the dramatic improvement in reducing the defect density, but there exists a more fundamental explanation for the performance gain after a critical
time. Prior studies suggest that an evolution time, or QSL, exists for certain Hamiltonians below which perfect state-to-state transfer cannot be guaranteed~\cite{Fleming,Bhattacharyya,Caneva,Bason,Hegerfeldt}.
Although the defect number $D$ is not related to fidelity in a simple way, we have observed that for unoptimized quenches the slowest mode $k_N$ dominates the sum $D=2\sum_{k>0} P_k$. Furthermore, the sharp drop in $\rho(T)$ seen in Fig.~\ref{defects} occurs \textit{concurrently} with a sharp drop in $P_{k_N}$. This can be seen in Fig.~\ref{Pk}. Once $T_c$ is passed, i.e., for $T=17.8>T_c$, the optimal power law quench results in a sharp drop of $P_{k_N}$, causing the drastic reduction of the overall defect density. This suggests that we may take maximization of the fidelity with the ground state in this mode as a surrogate for minimizing the defect density, allowing us to apply the QSL analysis to our model. We reason that the mode $k_N$ will have the highest QSL if one exists for each mode $k$. Since none of the modes must be excited in order to arrive at a defect-free ground state, the QSL of the entire system must be bounded below by the QSL of mode $k_N$.
\begin{figure}[h!]
  \includegraphics[width=.5\textwidth]{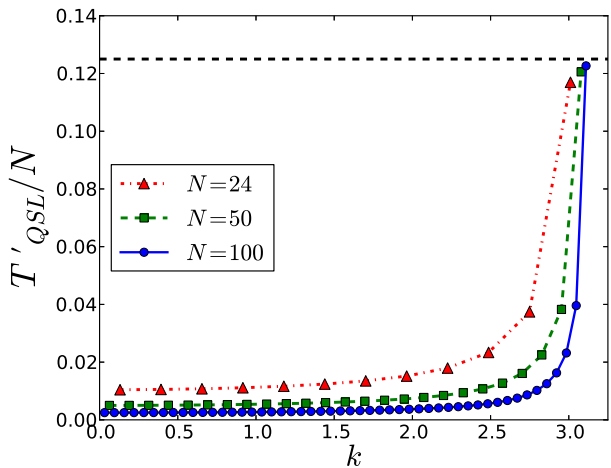}
  \caption{\label{qsl}The quantum speed limit $T'_{\rm{QSL}}(k)/N$ is plotted for three system sizes $N=24$, 50, and 100. The limiting value of $T'_{\rm{QSL}}(k_N)/N=1/8$ is plotted as the dashed line.}
\end{figure}
\par In the even subspace, the Hamiltonian $H_k$ of mode $k$ takes the Landau-Zener form $H^{(e)}_k=-\Gamma_k+\Gamma_k\sigma_z+\omega_k\sigma_x$ with $\Gamma_k=2(g+\cos k)$ and $\omega_k=-2\sin k$, which allows us to apply the results of Ref.~\cite{Hegerfeldt} to $H^{(e)}_{k_N}$, due to the fact that $\Gamma_{k_N}\approx 2(g-1)$ for $N\gg1$. For the symmetric quench we consider, the speed limit time for the mode $k_N$ is determined by $\tan[2\omega_{k_N} T_{\rm{QSL}}(k_N)]\propto 1/\omega_{k_N}$~\cite{Hegerfeldt}. Using $\frac{1}{\omega_{k_N}}\approx-N/(2\pi)\ll-1$, we have $2\omega_{k_N} T_{\rm{QSL}}(k_N)\approx -\pi/2$ or $T_{\rm{QSL}}(k_N)/N\approx1/8$, which is very close to our numerical values for $\tau_c$.
\par To get an estimate of the speed limit for all of the modes, we further observe that just above $\tau_c$, the optimal power $r^*$ takes high values, especially as $N$ increases. For large powers, the power law pulse is nearly constant at $g=1$ for much of the evolution. Therefore, as in Ref.~\cite{Caneva}, we apply the Fleming-Bhattacharyya bound~\cite{Fleming,Bhattacharyya} to the time-independent Hamiltonian $H^{(e)}_k(g=1)$, which gives $\cos [2\Delta E_kT'_{\rm{QSL}}(k)]=|\langle G_k(g_i)|G_k(g_f)\rangle|$. Here $\Delta E_k$ is the energy variance of the initial state $|G_k(g_i)\rangle$ with respect to $H^{(e)}_k(g=1)$.
Calculating $T'_{\rm{QSL}}(k_N)/N$ for $N=24$, 50, and 100 then yields $0.117$, $0.121$, and $0.123$, respectively, remarkably close to the numerically observed $\tau_c$ in Fig.~\ref{defects}. Since $\Delta E_{k_N}\approx|\omega_{k_N}|$, $T_{\rm{QSL}}(k_N)\approx T'_{\rm{QSL}}(k_N)$~\cite{Hegerfeldt}. Further support for taking $T'_{\rm{QSL}}(k_N)$ as the QSL for the entire chain comes from calculating $T'_{\rm{QSL}}(k)$ for every mode $k$, and observing that it falls off steeply as $k$ is decreased from $k_N$, as shown in Fig.~\ref{qsl}.
In the large $N$ limit, the approximation of a constant Hamiltonian improves, since $r^*$ becomes large, and the conjecture $T_c\approx T_{\rm{QSL}}(k_N)$ gains validity, as $T'_{\rm{QSL}}(k)$ falls off increasingly sharply from $k_N$ with increasing $N$ (Fig.~\ref{qsl}). Thus, we conjecture that a critical value of $\tau_c=1/8$ should be obtained in the limit $N\to\infty$.

The procedure above also allows for unrestricted optimization of the pulse shape. Starting from the optimal power law pulse, similar calculations were carried out without restricting the pulse shape to a power law, and we observed that only minor improvements resulted in both the optimized defect density and the observed value of $\tau_c$ (data not shown). These pulses are also quite robust, as moderate amounts of noise do not destroy the overall power law structure~\cite{SM}.

\par  To experimentally test our method, we note that the onset of a quantum phase transition from the paramagnetic to the ferromagnetic phase of a finite size Ising chain has been observed in trapped ion quantum simulators~\cite{NC,ionsreview}, most recently using up to $N=300$ spins~\cite{Britton}. An alternative platform would be neutral atoms loaded in an optical lattice, where nearest-neighbor Ising Hamiltonians have been successfully simulated~\cite{OP}. Our work is therefore germane to experimental and theoretical studies of non-equilibrium dynamics, quantum computation, and optimal manipulation of topological states of matter~\cite{MF,MFcontrol}, all of which require the precise preparation of initial states.

\par In conclusion, we have proposed a general gradient-based deterministic optimal control technique that can be used on a general class of critical systems. It is also straightforward to combine our method with the quantum driving approaches to achieve further suppression of the density of defects.

\emph{Acknowledgements}.--We would like to thank T.-S. Ho, G. Cohen, Y. Bar Lev, and D. Underwood for useful discussions. We acknowledge support from NSF Grant No. CHE-1058644 and ARO-MURI Grant No. W911NF-11-1-2068. The calculations in this work were performed at the TIGRESS computer center at Princeton University.

\setcounter{equation}{0}
\setcounter{figure}{0}
\section{Supplemental materials}
\subsection{A: Derivation of the gradient function Eq.~(\ref{gradO})}
We derive the gradient Eq.~(\ref{gradO}) for the free fermion model Eq.~(\ref{Hkmode}). For simplicity, we use $U(t)$ and $|G_i\rangle$ to denote the evolution operator $U(t,-T)$ and the initial ground state $|G(g_i)\rangle$, respectively. The functional derivative of the expectation value $O(T)$ of a general observable $\hat{O}$ with respect to the control field $g(t)$ can be calculated as~\cite{Brif}
\begin{eqnarray}
\frac{\delta O(T)}{\delta g(t)}=2\Im\langle G_i| \hat{O} (T)\hat{Y}(t)|G_i\rangle.
\end{eqnarray}
where $\hat{X}(t)\equiv U^\dag(t)\hat{X}U(t)$ is the Heisenberg picture operator for any operator $\hat{X}$. By assumption, $\hat{O}(T)$ and $\hat{Y}(t)$ can be written as $\hat{O}(T)=\sum_{\mathbf{k}}\hat{O}_{\mathbf{k}}(T),~\hat{Y}(t)=\sum_{\mathbf{k}}\hat{F}_{\mathbf{k} }(t)$, where we have used $U(t)=\prod_{\mathbf{k}}U_{\mathbf{k}}(t)$ and $X_{\mathbf{k}}(t)\equiv U_{\mathbf{k}}^\dag(t)X_{\mathbf{k}}U_{\mathbf{k}}(t)$. Noting that $|G_i\rangle=\prod_{\mathbf{k}}|G_{i,\mathbf{k}} \rangle$, then
\begin{eqnarray}
&&\frac{\delta O( T)}{\delta g( t)}=2\Im\sum_{\mathbf{k},\mathbf{k}'}\nonumber\\
 &&\prod_{\mathbf{p}(\neq \mathbf{k})}\langle G_{i,\mathbf{p}} |\langle G_{i,\mathbf{k}}| \hat{O}_\mathbf{k}(T) \hat{F}_{\mathbf{k}'} (t)|G_{i,\mathbf{k}'}\rangle\prod_{\mathbf{p}'(\neq \mathbf{k}')}|G_{i,\mathbf{p}'}\rangle
\end{eqnarray}
Now we separate the summation over $\mathbf{k}$ and $\mathbf{k}'$ for the two cases $\mathbf{k}\neq \mathbf{k}'$ and $\mathbf{k}=\mathbf{k}'$, so that
\begin{eqnarray}\label{grad3}
 \frac{\delta O( T)}{\delta g( t)}&=&2\Im\sum_{\mathbf{k}\neq \mathbf{k}'} \langle G_{i,\mathbf{k}}| \hat{O}_\mathbf{k}(T)|G_{i,\mathbf{k}}\rangle\langle G_{i,\mathbf{k}'} | \hat{F}_{\mathbf{k}'} (t)|G_{i,\mathbf{k}'} \rangle\nonumber\\
&&+2\Im\sum _\mathbf{k}\langle G _{i,\mathbf{k}} | \hat{O}_\mathbf{k}  (T)  \hat{F}_{\mathbf{k} } (t)|G_{i,\mathbf{k} } \rangle\nonumber\\
&= &2\Im [ \sum_\mathbf{k}\langle G _{i,\mathbf{k}} | \hat{O}_\mathbf{k}(T)  |G_{i,\mathbf{k}} \rangle][ \sum_{\mathbf{k}'}\langle G_{i,\mathbf{k}'} | \hat{F}_{\mathbf{k}'} (t)|G_{i,\mathbf{k}'} \rangle]\nonumber\\
&&-2\Im\sum_\mathbf{k}\langle G_{i,\mathbf{k}} | \hat{O}_\mathbf{k}  (T)|G_{i,\mathbf{k} } \rangle \langle G_{i,\mathbf{k} }|\hat{F}_{\mathbf{k} }(t)|G_{i,\mathbf{k} } \rangle \nonumber\\
&&+2\Im\sum _\mathbf{k}\langle G _{i,\mathbf{k}} | \hat{O}_\mathbf{k}  (T)  \hat{F}_{\mathbf{k} } (t)|G_{i,\mathbf{k} } \rangle\nonumber\\
&= &2\Im\sum _\mathbf{k}\langle G _{i,\mathbf{k}} | \hat{O}_\mathbf{k}  (T)|\bar{G}_{i,\mathbf{k} } \rangle \langle \bar{G}_{i,\mathbf{k} }|  \hat{F}_{\mathbf{k} } (t)|G_{i,\mathbf{k} } \rangle\nonumber\\
&= &2\Im\sum _\mathbf{k}\langle \phi_{ \mathbf{k}}(T) | \hat{O}_\mathbf{k} |\bar{\phi}_{ \mathbf{k}}(T)\rangle \langle \bar{\phi}_{ \mathbf{k} }(t)|  \hat{F}_{\mathbf{k} } |\phi_{\mathbf{k} }(t) \rangle.
\end{eqnarray}
In deriving the penultimate line of Eq. (\ref{grad3}) we have used the identity $1_\mathbf{k}=|G_{i,\mathbf{k}}\rangle\langle G_{i,\mathbf{k}}|+|\bar{G}_{i,\mathbf{k}}\rangle\langle \bar{G}_{i,\mathbf{k}}|$ and the Hermitian properties of $\hat{O}_\mathbf{k} $ and $\hat{F}_\mathbf{k}$.
\subsection{B: Robustness to noise}
\par
\begin{figure}[h!]
  \includegraphics[width=.5\textwidth]{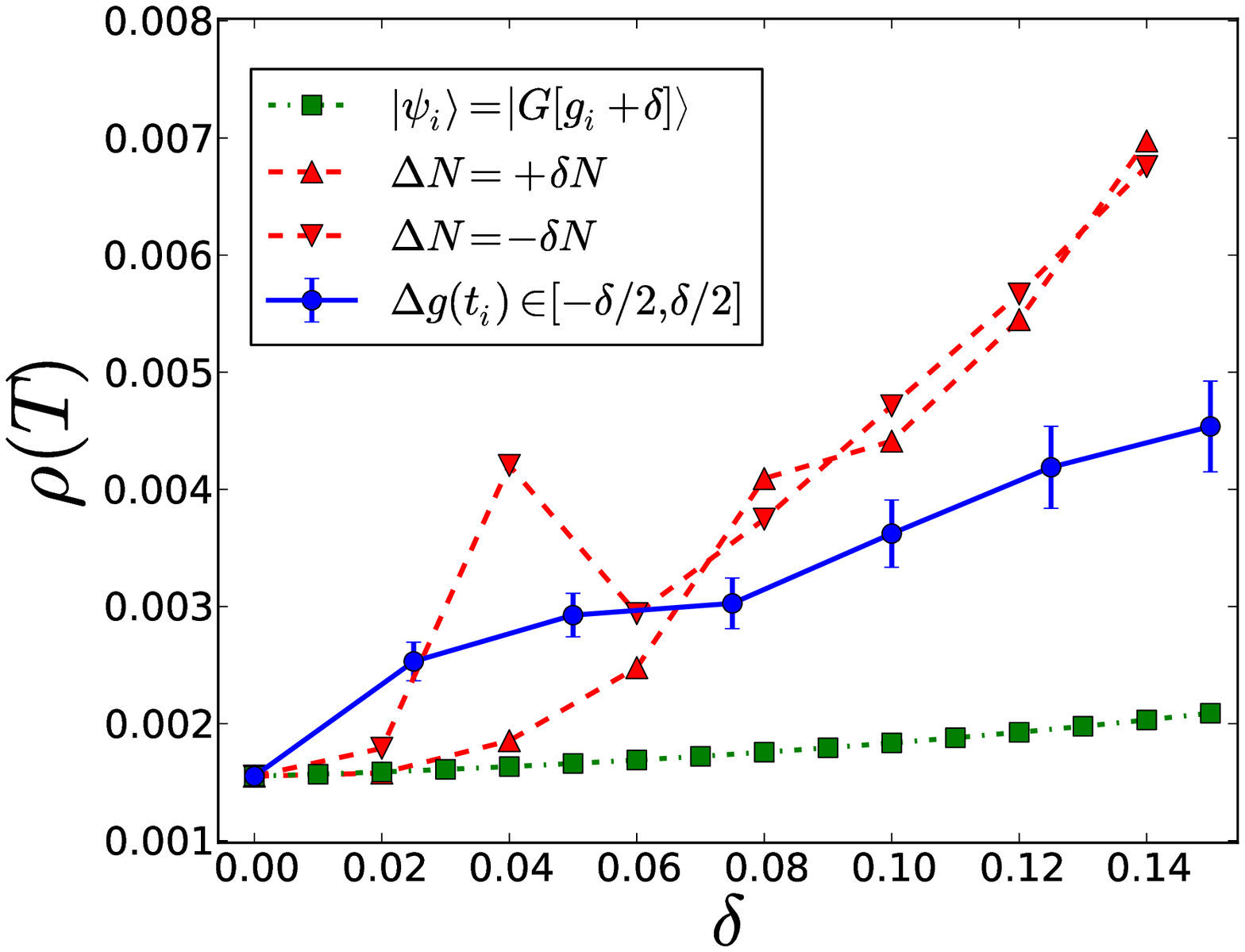}
  \caption{\label{noise} The average defect density resulting from applying pulses with random dynamical fluctuations characterized by strength $\delta$ (blue circles), using imprecisely prepared initial states (green squares), and using an incorrect number of spins (red triangles).
  The optimized power law pulse for $T=17.8$ and $N=100$ spins is used. The error bars represent $95\%$ confidence intervals.}
\end{figure}

 In this appendix, we will test the robustness of the obtained optimal power-law quench by using the gradient algorithm. We have checked this by adding random numbers, drawn from a uniform distribution $[-\delta/2,\delta/2]$, to the optimal power pulse at each time point for $N=100$ and $T=17.8$. For $\delta\leq 0.15$, after averaging over 500 realizations of noise, the final defect density does not increase by more than an order of magnitude (see Fig.~\ref{noise}). Also shown are the defect densities when the optimal power law pulse is used with incorrectly prepared initial states $\ket{G[g_i+\delta]}$ and with fluctuations in the number of spins $\Delta N=\pm N\delta$. Similar robustness is seen in these cases.

\end{document}